\documentclass[sigconf]{acmart}
\AtBeginDocument{%
  }

\usepackage{amsmath,amssymb,amsfonts}
\usepackage{algorithm2e}
\usepackage{graphicx}
\usepackage{circledsteps}
\usepackage{textcomp}
\usepackage{subfigure}
\usepackage{subcaption}
\usepackage{booktabs}
\usepackage{url}
\usepackage{colortbl}
\usepackage{xcolor}
\usepackage{dsfont}

\copyrightyear{2026}
\acmYear{2026}
\begin{document}
\title{From Servers to Sites: Compositional Power Trace Generation of LLM Inference for Infrastructure Planning}

\author{Grant Wilkins}
\email{gfw@stanford.edu}
\affiliation{%
  \institution{Stanford University}
  \city{Stanford}
  \state{CA}
  \country{USA}
}

\author{Fiodar Kazhamiaka}
\email{fkazhamiaka@microsoft.com}
\affiliation{%
  \institution{Microsoft Azure Research}
  \city{Redmond}
  \state{WA}
  \country{USA}
}

\author{Ram Rajagopal}
\email{ramr@stanford.edu}
\affiliation{%
  \institution{Stanford University}
  \city{Stanford}
  \state{CA}
  \country{USA}
}

\renewcommand{\shortauthors}{Wilkins et al.}


\newcommand{\fiodar}[1]{\textcolor{red}{#1}}

\begin{abstract}

Datacenter operators and electrical utilities rely on power traces at different spatiotemporal scales. Operators use fine-grained traces for provisioning, facility management, and scheduling, while utilities use site-level load profiles for capacity and interconnection planning. Existing datacenter power models do not capture LLM inference workloads, in which GPUs shift rapidly among compute-intensive prefill, lower-power decode, and idle states, and facility demand depends on how these states evolve and synchronize across many devices. We show that LLM inference power can be represented compositionally through two components: workload-driven transitions among operating states and configuration-specific power distributions within those states. Building on this observation, we develop a trace-generation framework that learns from measured traces and synthesizes power profiles for new traffic conditions and serving configurations. These traces aggregate from GPU servers to rack-, row-, and facility-scale load profiles at the temporal granularity required by the study.

Across multiple LLMs, tensor-parallel settings, and GPU generations, our framework achieves median absolute energy error below 5\% for most configurations while preserving temporal autocorrelation structure. The resulting traces support downstream analyses including oversubscription, power modulation, and utility-facing load characterization, enabling infrastructure evaluations that flat nameplate assumptions and static trace replay cannot support.
\end{abstract}

\maketitle

\section{Introduction}
\label{sec:introduction}
\begin{figure}
    \centering
    \includegraphics[width=\columnwidth]{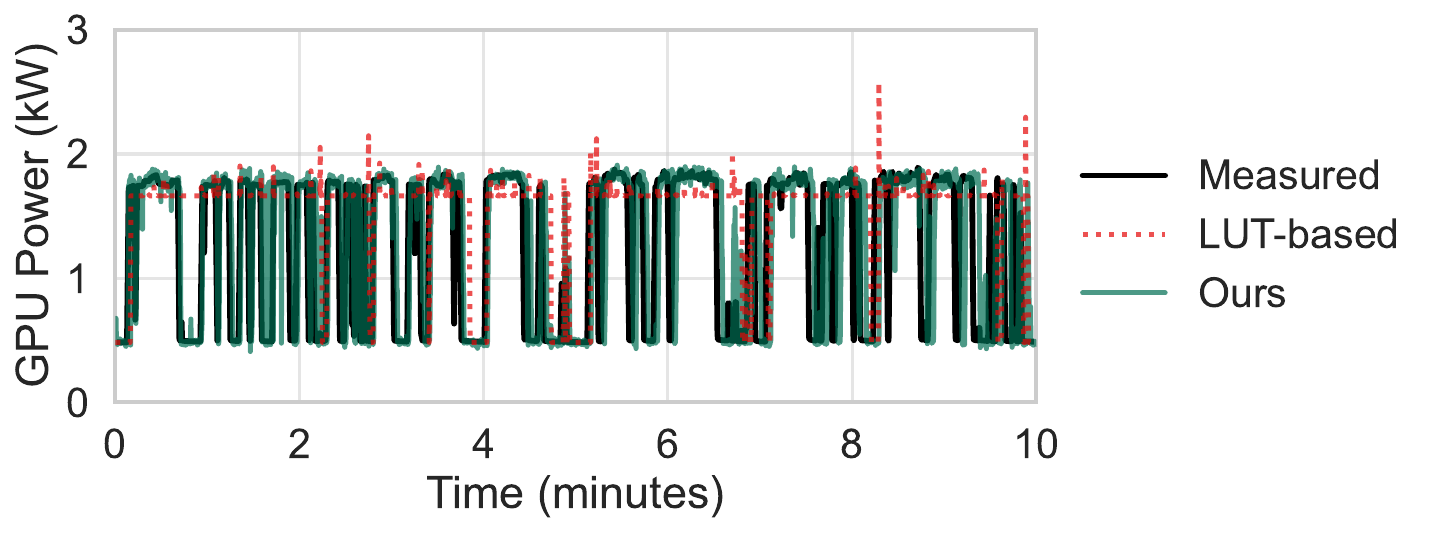}
    \caption{Server-level power trace comparison for Llama-3.1 (70B) with TP=8 on an A100 DGX. A phase-based look-up-table (LUT) baseline produces jumps and misses intermediate operating regions, ignoring the mixed prefill--decode occupancy from continuous batching. Our model more closely follows the measured trace across transitions between idle, partially loaded, and saturated regimes.}
    \label{fig:trace-comparison}
\end{figure}

Models of datacenter electrical demand inform decisions ranging from infrastructure provisioning inside the datacenter to interconnection planning on the grid. LLM inference has changed the character of that demand. Unlike conventional CPU-dominated services, modern inference deployments run on high-power GPU servers, often across multiple GPUs or multiple servers, and can drive large power swings on sub-second timescales as requests arrive and complete. As a result, facility-scale load can no longer be represented adequately by coarse utilization or average-power abstractions alone.

At the device level, LLM serving repeatedly moves GPUs between compute-heavy prefill, lower-power decode, and idle states~\cite{vllm,patel2024splitwise}. These transitions accumulate across a deployment, so facility demand depends not only on average utilization but also on how load shifts propagate and align across many GPUs. Capturing those effects requires modeling at server or GPU-device granularity, since the timing and synchronization of these transitions shape the aggregate load seen at larger scales~\cite{radovanovic2022modeling,powernap,pelley2010power,joulemeter,dynamollm}.

Datacenter operation and grid planning have different requirements for these load models. The former use power traces to study provisioning, oversubscription, power capping, and scheduling under changing hardware, model mixes, and traffic conditions~\cite{harvesting, muserve, maliakel2026characterizingllminferenceenergyperformance, radovanovic2022modeling, wilkins2024hybrid, li2019capmaestro}. The latter instead consume aggregate load profiles for interconnection and capacity studies, where conclusions depend on quantities such as peak demand, ramping behavior, and coincidence with other loads in the network~\cite{nerc2025largeloads, chen2025electricitydemandgridimpacts}. In both settings, the quality of the larger-scale model depends on whether it preserves inference-driven device-level dynamics that can generate those system-level characteristics.

Prior work captures important pieces of LLM inference power modeling~\cite{patel2024splitwise,dynamollm,stojkovic2025tapas,polca}, but not the combination of properties needed here: generalization to unseen traffic conditions, transfer across serving configurations, and aggregation into facility-scale load profiles for deployments that may not yet exist. Phase-based analyses identify recurring regimes such as idle, prefill, and decode~\cite{patel2024splitwise}, but fixed per-phase power levels cannot capture how power varies with offered load and concurrent occupancy. As Figure~\ref{fig:trace-comparison} shows, a phase-based look-up-table (LUT) baseline produces artificial jumps and misses intermediate operating regions created by overlapping prefill and decode work under continuous batching. Replay from measured deployments preserves a specific observed system~\cite{polca, stojkovic2025tapas, dynamollm}, but does not extrapolate reliably to new arrival processes, model mixes, or deployment scales. What is missing is a representation that captures both the workload dynamics governing transitions among operating states and the configuration dependence of power within each state.

Our central insight is that inference power traces can be generated compositionally. Workload dynamics determine the evolution of operating state over time, while model choice, hardware platform, and serving configuration determine the distribution of power conditional on that state. This representation transfers across both traffic scenarios and deployment settings.

\begin{figure}[!htb]
    \centering
    \includegraphics[width=\linewidth]{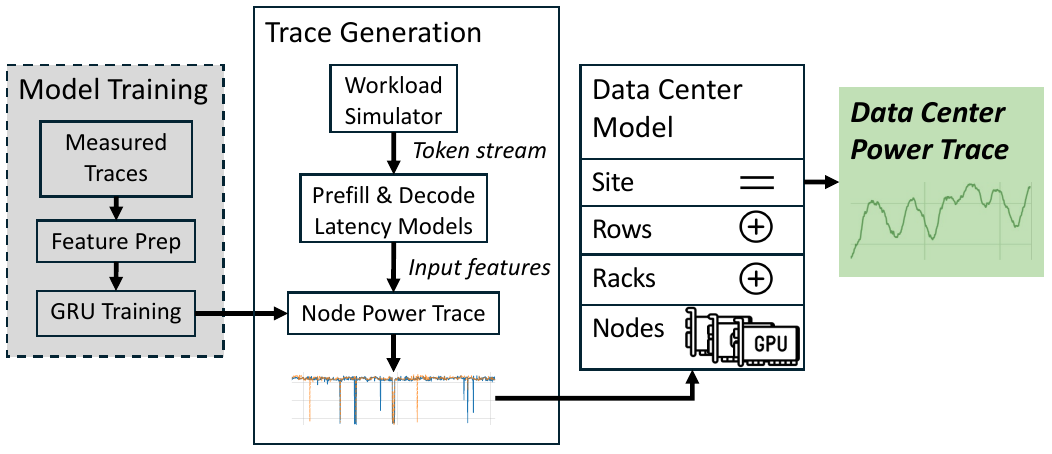}
    \caption{Compositional trace generation pipeline. Measured power traces train per-configuration state classifiers and state-conditioned power models offline. At generation time, a workload simulator produces token-level features from an arrival schedule; the classifier maps features to states; the power model converts the resulting state trajectory into server-level power traces; and these traces aggregate through the datacenter hierarchy to produce facility-scale load profiles.}
    \label{fig:overall-figure}
\end{figure}

Figure~\ref{fig:overall-figure} illustrates the framework. Measured traces are used offline to train a classifier that predicts transitions among operating states from workload features. At generation time, a simulator converts a stream of request arrivals into a feature sequence, the classifier produces a state trajectory, and a configuration-specific power model maps that trajectory to a power trace. This allows us to synthesize traces for new settings without remeasuring each an end-to-end scenario, and to aggregate them from GPU servers to rack-, row-, and facility-scale profiles at the temporal resolution required by a study.

Our contributions are threefold. We present a compositional model for generating LLM inference power traces that supports both fine-grained operational analysis and planner-facing facility studies. We show that this separation creates a transferable representation across arrival patterns, hardware platforms, and model mixes. Finally, through case studies, we demonstrate that inference-aware traces can materially change conclusions relative to TDP-based and phase-only abstractions.

We evaluate this pipeline on five dense and two mixture-of-experts (MoE) LLMs across A100 and H100 GPUs and multiple tensor-parallel settings. For most dense-model settings, the generator achieves median energy error below 5\% while preserving temporal structure with high autocorrelation fidelity. We then combine the framework with production Azure token-traffic traces to construct datacenter power traces that can be used in downstream planning studies. Relative to TDP-based provisioning and simpler phase-based models, our traces produce materially different estimates of peak demand, ramping, and feasible oversubscription levels, showing that trace fidelity is not merely descriptive but can change infrastructure conclusions. We release the data-collection scripts, trained classifiers, and trace-generation code to support reproducibility and downstream use\footnote{\url{https://github.com/grantwilkins/powertrace-sim}}.

\section{Motivation and Observations}
\label{sec:background}

\subsection{Power States in Continuous-Batch Inference}
\label{ssec:phase-structure}

Modern LLM inference engines use continuous batching~\cite{vllm, patel2024splitwise} to serve requests across a set of GPUs. When a new request arrives, its prompt is processed in a compute-bound \textit{prefill} pass that drives GPU power to 80--90\% of TDP~\cite{patel2024splitwise, stojkovic2025tapas, prefillonly}. The request then enters a memory-bound \textit{decode} phase that generates tokens autoregressively at 40--60\% of TDP~\cite{mughees2025shorttermloadforecastingaidata, dynamollm}. Under continuous batching, prefill and decode often coexist within the same batch, so GPU power is determined not by a single phase label but by the \textit{mixture} of work executing at that moment~\cite{shuffleinfer}.

\begin{figure}
    \centering
    \includegraphics[width=\linewidth]{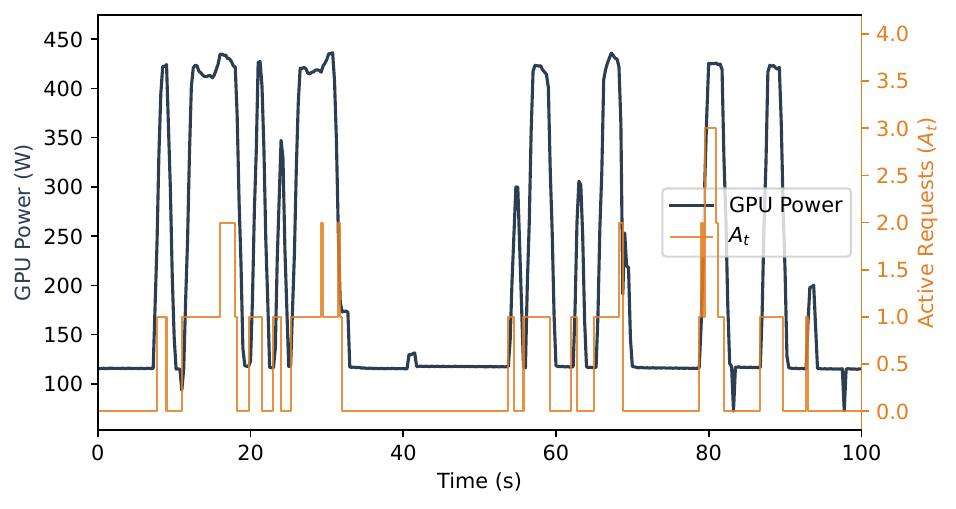}
    \caption{Measured GPU power and active request count $A_t$ for Llama-3.1-8B on H100 at $\lambda = 0.25$\,req/s. Each power transition aligns with a change in $A_t$, as sharp rises correspond to request arrivals (prefill), sustained plateaus to concurrent decode, and drops to request completions. The two signals move together throughout the trace.}
    \label{fig:at-overlay}
\end{figure}

This implies that GPU power is largely a function of the number of concurrently active requests and whether prompt-processing work is present in the batch. Figure~\ref{fig:at-overlay} makes this relationship visible, since measured GPU power and the active request count $A_t$ move closely together, and each major power transition aligns with a change in $A_t$. Sharp increases coincide with new request arrivals that trigger prefill, sustained elevated regions reflect concurrent decode activity, and drops toward idle occur as requests complete.

We formalize this observation using two features derived from the request schedule. Let $A_t$ denote the number of concurrently active requests at timestep $t$, and let $\Delta A_t = A_t - A_{t-1}$. A request is \textit{active} from the timestep its prefill begins until the timestep its final token is generated. The feature $A_t$ captures instantaneous load, as each active request occupies batch capacity, reserves KV cache, and contributes attention and feedforward work. As illustrated in Figure~\ref{fig:at-overlay}, total power scales with $A_t$ up to a hardware-dependent saturation point. The feature $\Delta A_t$ serves as a proxy for prefill pressure, since under continuous-batching policies that process prompts on admission, $\Delta A_t > 0$ strongly indicates that prefill is occurring alongside decode.

Both features can be computed from a request schedule together with a simple throughput surrogate for continuous-batching behavior. The surrogate requires only a small number of serving parameters, including a token-throughput curve, maximum batch size, and prompt and output length distributions. Given an arrival process and length distributions, one can therefore compute $A_t$ and $\Delta A_t$ for hypothetical deployments without coupling the power model to a specific serving-system implementation.

These features are useful not only because they explain server-level power variation, but also because they preserve the aspects of trace structure that downstream operational and planning workflows consume.

\subsection{Why Trace Fidelity Matters}
\label{ssec:planning-workflows}

The same underlying inference workload feeds multiple infrastructure workflows, each of which requires a different view of power.
For datacenter operators, the key input is a high-resolution IT-power trace that preserves transients, burst durations, and coincidence across nearby servers. These traces inform at least three classes of decisions~\cite{radovanovic2022modeling,li2019capmaestro}. First, sub-second control loops such as power capping, throttling, and protection logic must respond to excursions that threaten local power-delivery limits or protection thresholds~\cite{choukse2025powerstabilizationaitraining, li2024unseenaidisruptionspower, stojkovic2025tapas, spaan2026reducingcomputewastellms,li2019capmaestro}. Second, over timescales of seconds to minutes, operators study whether excess electrical capacity can be traded for additional IT load through oversubscription, placement, or load-balancing policies~\cite{polca, harvesting, smartoclock, hsu2018smoothoperator, stojkovic2025tapas, zhang2021flex,mvplane-powercapping}. Third, over planning horizons of months to years, fleet designers need scenario-conditioned traces to estimate how row- and site-level demand changes as accelerator generations, model mixes, and traffic volumes evolve~\cite{lin2024exploding, baxi2025online, tcopaper}.

For these workflows, usable headroom depends on whether traces preserve peak magnitude, peak duration, coincidence across nearby servers, and how bursts change under different batching and traffic conditions. A \textit{flat TDP} assumption is conservative but often overly restrictive, whereas a \textit{smoothed mean} power profile can be misleading in the opposite direction because it suppresses short excursions.

Grid-facing studies consume a coarser version of this same trace. Utility workflows typically require metered load shapes together with statistics such as interval peaks, ramp-rate distributions, coincidence with surrounding demand, and peak duration at the billing or protection timescale~\cite{EPRI2024PoweringDataCenters, capacityplanning}. In some jurisdictions, projected load shapes are now part of the datacenter connection process itself~\cite{AESO2025datacentreconnection}. Similar inputs also appear in capacity-expansion and resource-adequacy studies, where the question is whether generation and network infrastructure can accommodate a large new load under simulated interactions with the rest of the system~\cite{gridstrategies2025, chen2025electricitydemandgridimpacts,nerc2025largeloads,nerc2026reliability}. For these studies, the key issue is how inference activity aggregates to the point of common coupling (PCC): i.e., whether short bursts wash out under diversity, persist into metered peaks, or align with an already-stressed system. In our setting, facility power at the PCC is derived from modeled IT power using the datacenter aggregation procedure described in \S\ref{ssec:datacenter}. 

\section{Methods}
\label{sec:methods}

\subsection{Planner-Facing Interface}
\label{ssec:interface}

The framework exposes a planner-facing interface: given a facility configuration and workload scenario, it returns a power time series at user-specified temporal resolution. The required inputs are:
\begin{itemize}
    \item \textbf{Facility topology:} number of rows, racks per row, and servers per rack.
    \item \textbf{Server configuration:} GPU type (e.g., A100 or H100), model family and size, and tensor-parallel degree.
    \item \textbf{Workload scenario:} request arrival process (e.g., Poisson with rate $\lambda$) together with prompt- and output-length distributions.
    \item \textbf{Site-level assumptions:} per-server non-GPU IT power and site power usage effectiveness (PUE).
\end{itemize}
The output is a per-server power trace at 250\,ms resolution. These traces can be aggregated to rack, row, and site levels and resampled at any coarser interval needed by a downstream study. Site-level output represents IT load scaled by PUE. Any onsite storage or generation must be modeled externally. Internally, the pipeline consists of three stages: (1) state modeling (\S\ref{ssec:model}), (2) trace synthesis (\S\ref{ssec:generation}), and (3) datacenter-scale aggregation (\S\ref{ssec:datacenter}).

\subsection{Power State Modeling}
\label{ssec:model}

\begin{figure}
    \centering
    \includegraphics[width=\linewidth]{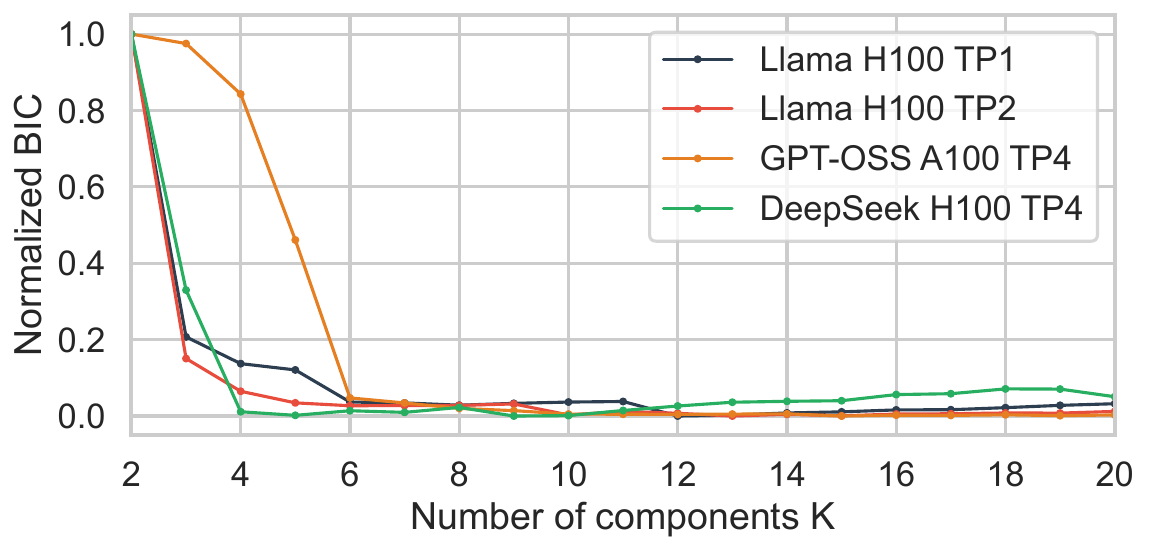}
    \caption{Normalized BIC as a function of mixture components $K$ for four representative configurations. BIC plateaus near $K{=}10$ for most configurations, with selected $K$ ranging from 8 to 12.}
    \label{fig:bic-selection}
\end{figure}

\noindent\textbf{State discovery via Gaussian mixtures.}
Measured power traces show that GPU power during LLM inference concentrates in a small number of recurring operating regimes. Rather than fixing these regimes a priori, we infer them separately for each hardware--model--deployment\footnote{In our study we vary deployment with tensor parallelism (TP); however, this applies for other variations in serving policy as well.} configuration.

For each configuration $(H, M, \mathrm{TP})$, we fit a Gaussian mixture model with $K$ components:
\begin{equation}
    p(y_t \mid H, M, \mathrm{TP}) = \sum_{k=1}^{K} \pi_k \mathcal{N} \left(y_t \mid \mu_k, \sigma_k^2\right),
    \label{eq:gmm}
\end{equation}
where $y_t$ is measured GPU power at time $t$, $\pi_k$ is the mixing weight of component $k$, and $(\mu_k, \sigma_k^2)$ are its mean and variance. We choose $K$ independently for each configuration using the Bayesian Information Criterion (BIC). As shown in Figure~\ref{fig:bic-selection}, BIC typically plateaus near $K=10$, and the selected values range from 8 to 12 across our measured configurations.

To train our temporal model, we take each measured power sample as a hard state label from posterior maximization:
\begin{equation}
    z_t = \arg\max_k \pi_k \mathcal{N} \left(y_t \mid \mu_k, \sigma_k^2\right),
    \label{eq:state-assignment}
\end{equation}
and sort components by mean power to obtain an ordered set of operating states ranging from idle to full-load behavior. Dense models typically yield well-separated states, while MoE models exhibit broader and more overlapping components. The resulting state dictionary,
$\{(\mu_k^{(H,M,\mathrm{TP})}, \sigma_k^{(H,M,\mathrm{TP})})\}_{k=1}^K$,
serves both as the labels for temporal classification and as the power sampling model used during trace generation. Because Gaussian components are only an empirical approximation to bounded hardware power, generated samples are clipped to the observed power range of the corresponding configuration.

\noindent\textbf{Temporal state classification.}
The GMM yields latent states for measured traces, but trace generation requires the inverse mapping, from workload features to state sequence. As motivated in \S\ref{ssec:phase-structure}, we describe the workload at each timestep using the active-request count $A_t$ and its first difference $\Delta A_t$ as features.

We model the mapping from workload features to latent states with a bidirectional GRU classifier. Let $\mathbf{X} = (\mathbf{x}_1,\ldots,\mathbf{x}_T)$ denote the feature sequence, where $\mathbf{x}_t = (A_t, \Delta A_t) \in \mathbb{R}^2$. The classifier outputs
\begin{equation}
    \mathbb{P}(z_t = k \mid \mathbf{X}) = \mathrm{softmax}(W\mathbf{h}_t + \mathbf{b})_k,
    \label{eq:bigru-output}
\end{equation}
where $\mathbf{h}_t \in \mathbb{R}^{2d}$ is the concatenated forward and backward hidden state at time $t$, and $W \in \mathbb{R}^{K \times 2d}$ projects it to $K$ state logits. Because our setting is offline trace generation for planning, the full workload is available at evaluation time, hence a bidirectional model is appropriate.

Among several sequence models we evaluated, the BiGRU provided the best tradeoff between temporal classification quality and downstream energy fidelity on our dataset. For configurations with sharp state transitions, the classifier typically produces concentrated posteriors; for smoother occupancy ramps, probability mass is often shared across adjacent states, allowing for some ambiguity near transition boundaries.

\subsection{Trace Generation}
\label{ssec:generation}

Given a new workload scenario, trace generation proceeds in three stages: (i) compute workload features from the arrival schedule, (ii) predict a latent state trajectory, and (iii) sample power values conditioned on that trajectory.

\noindent\textbf{From arrival schedules to workload features.}
We begin with a request arrival schedule
$\{(t_i, n_i^{\mathrm{in}}, n_i^{\mathrm{out}})\}_i$,
where $t_i$ is the arrival time and $n_i^{\mathrm{in}}$ and $n_i^{\mathrm{out}}$ are the input and output token counts. To compute the workload features $A_t$ and $\Delta A_t$, we estimate each request's active interval $[t_i^{\mathrm{start}}, t_i^{\mathrm{end}}]$ using a lightweight throughput surrogate calibrated per configuration.

Our surrogate models query lifetime as the sum of a prefill phase and a decode phase. Prefill duration (time-to-first-token, TTFT) grows superlinearly~\cite{pope2023efficiently,stojkovic2025tapas} with prompt length:
\begin{equation}
    \log(\mathrm{TTFT}) = \alpha_0 + \alpha_1 \log(n_{\mathrm{in}} + 1) + \varepsilon,
    \qquad \varepsilon \sim \mathcal{N}(0,\sigma^2_{\mathrm{TTFT}}),
    \label{eq:ttft}
\end{equation}
while decode duration is the product of output length and inter-token latency~\cite{vllm,stojkovic2025tapas} (time-between-tokens, TBT),
\begin{equation}
    \log(\mathrm{TBT}) \sim \mathcal{N}(\mu_{\log \mathrm{TBT}}, \sigma^2_{\log \mathrm{TBT}}).
    \label{eq:tbt}
\end{equation}
The parameters
$(\alpha_0,\alpha_1,\sigma_{\mathrm{TTFT}},\mu_{\log \mathrm{TBT}},\sigma_{\log \mathrm{TBT}})$
are estimated per configuration from measured traces, but they can also be obtained from a small benchmark sweep or supplied directly from deployment SLOs/SLAs.

Requests are then placed into a FIFO queue with batch size 64. Request $i$ begins execution at
$t_i^{\mathrm{start}} = \max(t_i, \text{earliest available slot})$,
incurs TTFT for prefill, and then decodes for
$n_i^{\mathrm{out}} \times \mathrm{TBT}$ seconds. We can then efficiently compute our features as,
\begin{equation}
    A_t = \left| \{ i : t_i^{\mathrm{start}} \le t < t_i^{\mathrm{end}} \} \right|,
    \label{eq:at-def}
\end{equation}
and $\Delta A_t = A_t - A_{t-1}$.

\begin{figure}
    \subfigure[Prefill]{
        \label{fig:prefill}
        \includegraphics[width=0.48\columnwidth]{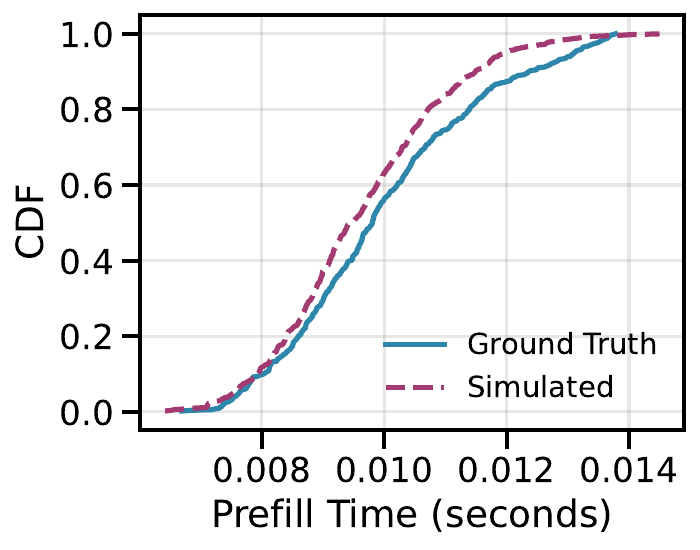}}
    \hfill
    \subfigure[Decode]{
        \label{fig:decode}
        \includegraphics[width=0.48\columnwidth]{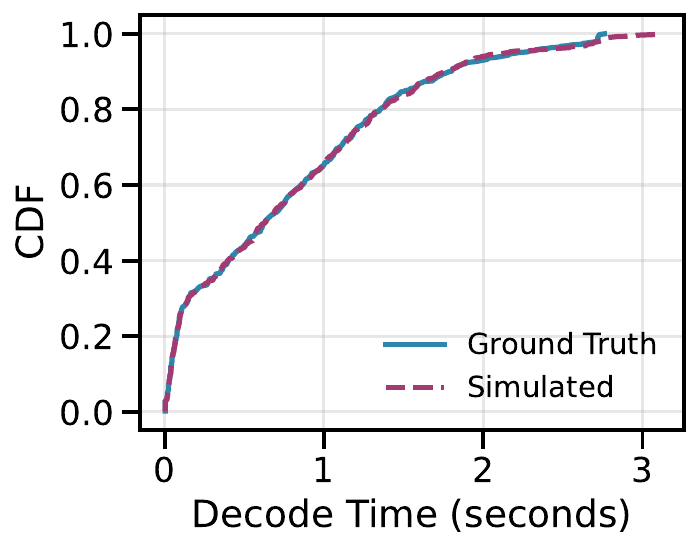}}
    \caption{CDFs of modeled versus measured prefill and decode durations for DeepSeek-R1-Distill (8B) on H100 with TP$=$8.}
    \label{fig:prefill-decode-cdf}
\end{figure}

This model is intended to reproduce request lifetimes and concurrency, not to emulate a scheduler fully. Different serving policies enter only through their aggregate effect on TTFT, TBT, and the resulting concurrency process. Figure~\ref{fig:prefill-decode-cdf} shows distributional agreement of this proxy against measured prefill and decode durations, and Appendix~\ref{app:surrogate} compares the resulting simulated and measured workload features.

\noindent\textbf{From workload features to a state trajectory.}
Given the feature sequence $\mathbf{X}'$, our trained BiGRU produces per-timestep state probabilities
$\mathbb{P}(z_t = k \mid \mathbf{X}')$.
We then can generate a state trajectory by sampling
\begin{equation}
    \hat{z}_t \sim \mathrm{Categorical}(\mathbf{p}_t),
    \qquad p_t(k) = \mathbb{P}(z_t = k \mid \mathbf{X}'),
    \label{eq:state-sampling}
\end{equation}
rather than taking an argmax at each timestep. At high loads, we typically remain within a single state. More uncertainty happens near transitions, where adjacent states may have similar probability.

\noindent\textbf{From state trajectory to power.}
Using the predicted state sequence $\{\hat{z}_t\}$, we sample power from the GMMs stored in the state dictionary. For dense transformer models, variation within a state is weakly correlated in time, so we can sample power independently from the corresponding GMM component:
\begin{equation}
    \hat{y}_t \sim \mathcal{N}\!\left(\mu_{\hat{z}_t}, \sigma^2_{\hat{z}_t}\right).
    \label{eq:power-iid}
\end{equation}

MoE models have stronger time correlated variation due to data-dependent expert routing. In this case, independent sampling can destroy autocorrelation that we observe but that is not recoverable from our workload features. We therefore use a per-state AR(1) model,
\begin{equation}
    \hat{y}_t = \mu_{\hat{z}_t}
    + \phi_{\hat{z}_t}(\hat{y}_{t-1} - \mu_{\hat{z}_t})
    + \sigma^{\mathrm{noise}}_{\hat{z}_t}\varepsilon_t,
    \qquad \varepsilon_t \sim \mathcal{N}(0,1),
    \label{eq:power-ar1}
\end{equation}
where $\phi_k$ is estimated from segments in the training data and
$\sigma_k^{\mathrm{noise}} = \sigma_k \sqrt{1-\phi_k^2}$
captures marginal variance of state $k$. Dense configurations typically have a $\phi_k$ near zero, whereas the MoE is higher. All generated samples are clipped to the observed power range $[y_{\min}, y_{\max}]$ from the training data.

\subsection{Datacenter-Scale Aggregation}
\label{ssec:datacenter}

\noindent\textbf{Facility hierarchy.}
We model each data hall as a four-level hierarchy~\cite{zhang2021flex}:
data hall $\rightarrow$ rows $\rightarrow$ racks $\rightarrow$ servers.
A hall with $s$ rows, $r$ racks per row, and $n$ servers per rack contains
$N_{\mathrm{total}} = s r n$
inference servers. Each server $N_{i,j,k}$ is assigned a configuration tuple
$(H, M, \mathrm{TP})$,
which selects a model. This framework lets us simulate arbitrary per-server assignments, and having heterogeneous mixes of accelerator generations, model sizes, and serving configurations within a single hall.

\noindent\textbf{Cross-server arrival structure.}
Aggregation depends not just on per-server power but also on how request streams are distributed across servers~\cite{hsu2018smoothoperator}. We support two primary traffic cases. First we have an independent case where each server draws from its own arrival process. Then we model a shared-intensity case in which servers follow a common arrival-rate function with independent thinning, yielding correlated request streams. This is the mode used with the Azure trace from~\cite{dynamollm}. These modes allow a planner to explore effects from independent to correlated traffic without needing to change the per-server trace-generation model. 

\noindent\textbf{Bottom-up aggregation.}
Let $y_t^{(i,j,k)}$ denote the generated \emph{GPU} power trace for server $N_{i,j,k}$. We model additional non-GPU IT load (e.g., CPUs, storage, and networking) as a constant
$P_{\mathrm{base}} = 1\,\mathrm{kW}$
per server. Total IT power is then
\begin{equation}
    P_{\mathrm{IT}}(t)
    =
    \sum_{i=1}^{s}\sum_{j=1}^{r}\sum_{k=1}^{n}
    \left(y_t^{(i,j,k)} + P_{\mathrm{base}}\right),
    \label{eq:it-power}
\end{equation}
from which rack, row, and hall level trajectories are just partial sums across time in the system. Our choice for a constant non-GPU term is motivated by the relative stability and magnitude difference of these components power draw during inference as compared to GPUs~\cite{warehouse, wilkins2024offline}.

\noindent\textbf{Facility power.}
We map IT power to facility power at the point of common coupling using a constant-PUE model,
\begin{equation}
    P_{\mathrm{facility}}(t) = \mathrm{PUE} \times P_{\mathrm{IT}}(t),
    \label{eq:facility-power}
\end{equation}
following standard datacenter planning practice~\cite{wu2016dynamo, radovanovic2022modeling,warehouse}. This approximation is appropriate for aggregate provisioning and grid-integration studies. We point out that our model does not capture weather-dependent shifts in PUE, as we discuss further in \S\ref{sec:discussion}.

\section{Evaluation}
\label{sec:evaluation}

We evaluate the framework across an increasing scale. We begin with server-level fidelity, showing that generated traces reproduce measured GPU power dynamics (\S\ref{ssec:node-fidelity}). We then compare against simpler baselines to identify which aspects of structure they miss (\S\ref{sec:baselines}). Next, we study whether those differences matter in a facility-scale case study driven by a production workload (\S\ref{ssec:facility}). Finally, we examine how aggregation through the datacenter hierarchy can smooth variability and the planning quantities derived from it (\S\ref{ssec:aggregation}).

\subsection{Experimental Setup}
\label{ssec:setup}

We evaluate our framework on Microsoft Azure NVIDIA DGX servers with $8\times$A100 (80\,GB) and $8\times$H100 (80\,GB) GPUs. All experiments use vLLM~\cite{vllm} v0.10.0 with default settings, except for tensor parallelism, which is varied to show robustness across different serving configurations.

GPU clocks and power caps are left unchanged from their default settings. Power is measured using \texttt{nvidia-smi} at 250\,ms intervals. This resolution is sufficient to resolve prefill and decode transitions, even though it does not capture sub-millisecond transients. Our measurement set spans 7 models, 2 GPU generations, and the tensor-parallel settings supported by each configuration.

The workload trace we validate our site-level activity on is a single-day snapshot (5/16/24) from a coding activity trace released by Microsoft~\cite{dynamollm}. The size of the facility, the size of models served, and the number of instances from this model were not publicly announced.

\noindent\textbf{Workload collection.}
For each configuration, we collect traces at 7 arrival rates ranging from 0.125 to 4\,req/s. Each trace contains $600\lambda$ prompts, corresponding to approximately 10 minutes of runtime, and we repeat each arrival rate 5 times. Request streams are drawn from four prompt datasets: ShareGPT, InstructCoder~\cite{instructcoder}, AIMO-Validation-AIME, and Edit-10K-Char.

We profile dense models from the Llama-3.1 family~\cite{llama3} at 8B, 70B, and 405B parameter counts. We also profile dense chain-of-thought reasoning models from the DeepSeek-R1~\cite{deepseek} distillations onto the 8B and 70B Llama-3.1 model architectures. Finally, we profile gpt-oss~\cite{openai2025gptoss120bgptoss20bmodel} 20B and 120B, two MoE models from OpenAI.

\noindent\textbf{Training.}
For each hardware, model, and parallelism configuration, we train a separate BiGRU classifier with hidden size $H=64$ and input dimension 2. The number of latent states $K$ is selected by BIC. Training data is split at the trace level into 70/15/15 train/validation/test partitions after pooling across arrival rates. All reported fidelity metrics are computed on held-out test traces.

\noindent\textbf{Metrics.}
We judge trace quality with four quantities. \emph{ACF~$R^2$} measures agreement between the autocorrelation function of measured and synthetic traces. \emph{$\Delta$Energy} measures signed relative error in total energy consumption, $\Delta E  = \left(E_{\mathrm{syn}} - E_{\mathrm{meas}}\right) / {E_{\mathrm{meas}}}$,
over the entire held-out trace. For each held-out trace, we generate 5 synthetic
traces using different random seeds and report the median $|\Delta E|$. \emph{KS statistic} measures whether distributionally our measured and synthetic power samples match. \emph{NRMSE} measures point-wise error normalized by the observed power range. For each held-out trace, we generate 5 synthetic traces using different random seeds and report the median metric value.

\subsection{Server-Level Trace Fidelity}
\label{ssec:node-fidelity}

\begin{figure*}[t]
    \subfigure[Dense model, sparse load ($\lambda = 0.25\,$req/s)]{
        \label{fig:trace-0.25}
        \includegraphics[width=0.48\textwidth]{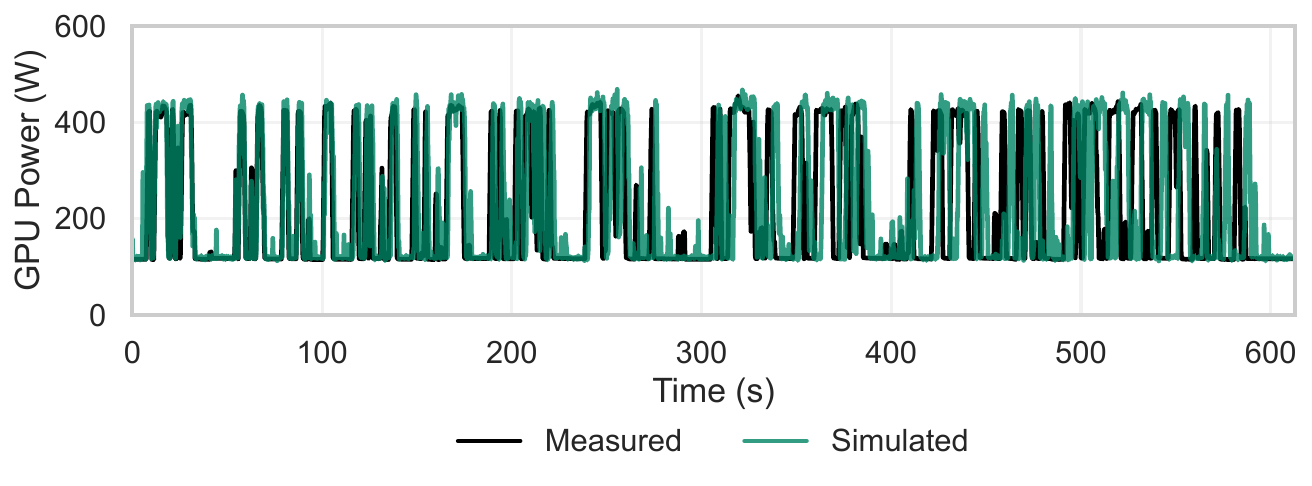}}
    \hfill
    \subfigure[Dense model, medium load ($\lambda = 1.0\,$req/s)]{
        \label{fig:trace-1.0}
        \includegraphics[width=0.48\textwidth]{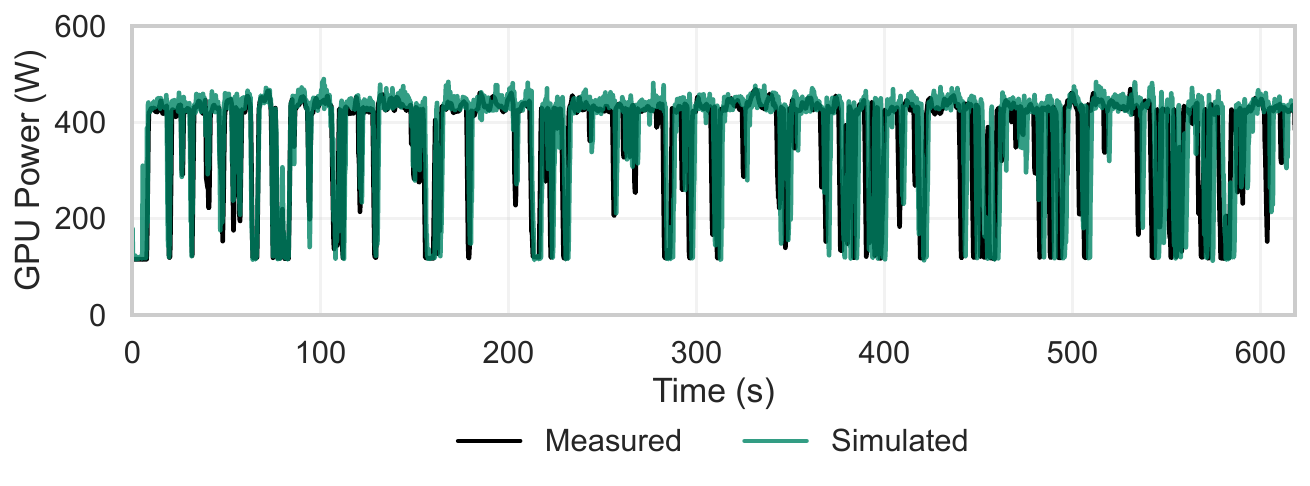}}
    \subfigure[Dense model, high load ($\lambda = 4.0\,$req/s)]{
        \label{fig:trace-4.0}
        \includegraphics[width=0.48\textwidth]{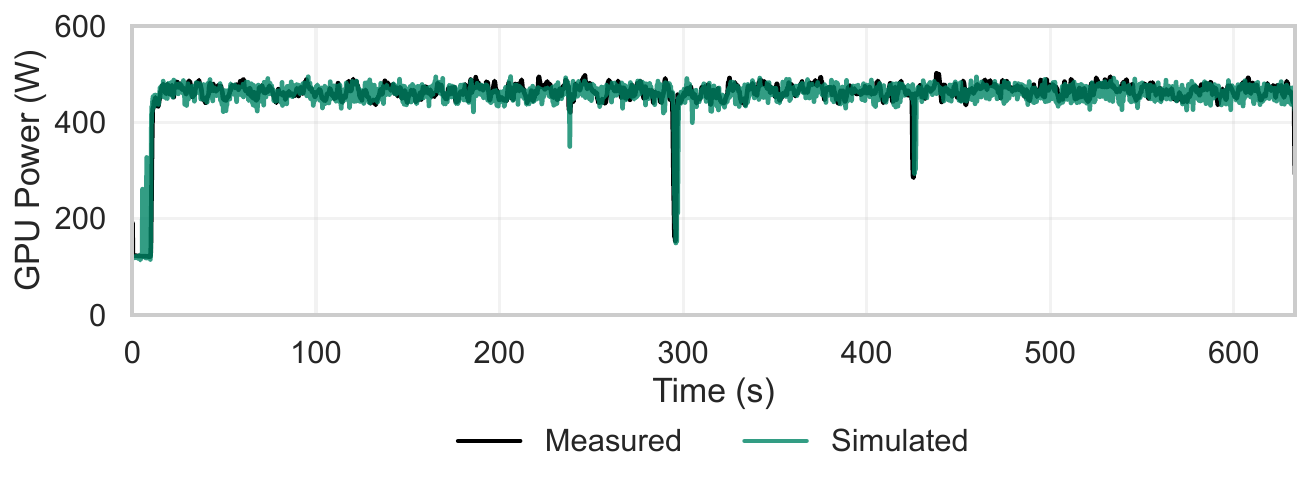}}
    \hfill
    \subfigure[MoE model, medium load ($\lambda = 1.0\,$req/s)]{
        \label{fig:trace-moe}
        \includegraphics[width=0.48\textwidth]{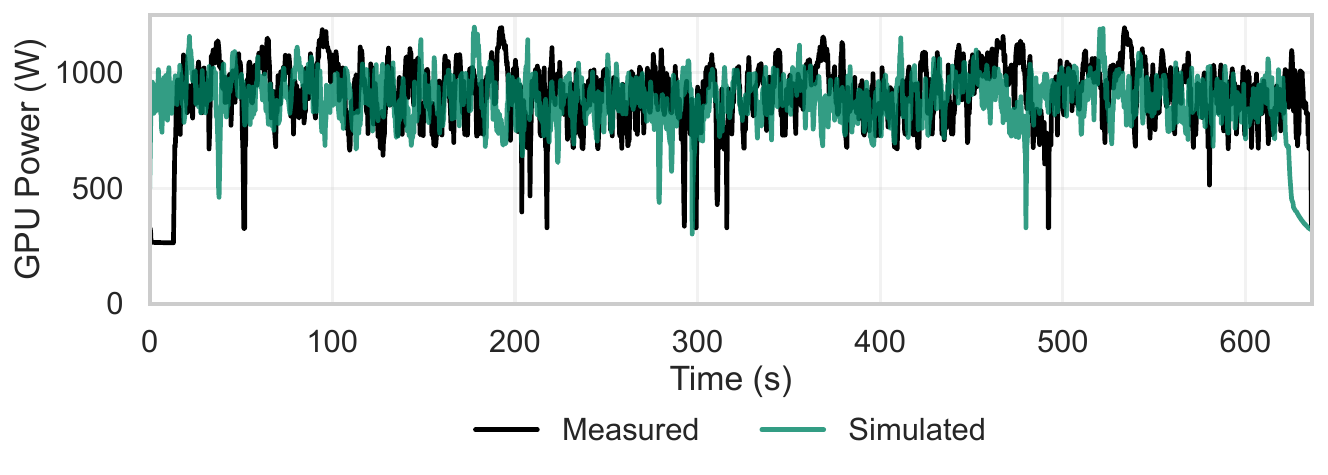}}
    \caption{Measured and simulated power traces for Llama-3.1 (8B) on A100 with TP$=$2 across three arrival rates (a--c), and for gpt-oss (120B) on A100 with TP=4 configuration (d).}
    \label{fig:time-series}
\end{figure*}

At the server level, our generated traces closely match measured dense-model power dynamics across a range of arrival rates, while MoE traces remain more challenging because expert routing causes within-state variation that is not observable from workload features alone. Figure~\ref{fig:time-series} shows representative examples.

For dense models at low load (Figure~\ref{fig:trace-0.25}), the measured trace alternates between idle and active levels as individual requests are served. The synthetic trace reproduces the timing and magnitude of these transitions. At moderate load (Figure~\ref{fig:trace-1.0}), overlapping requests have intermediate occupancy buckets they fall in, and the synthetic trace tracks these transitions with some minor smoothing near these boundaries. At high load (Figure~\ref{fig:trace-4.0}), the GPU operates at saturation, with some dips when the queue is empty.

The MoE example in Figure~\ref{fig:trace-moe} exhibits more within state variation and much more temporal persistence than dense traces. Our AR(1)-based sampling recovers some of this structure, giving some gradual within-state movement rather than the i.i.d.\ jitter that we would see in dense-model sampling, but the match is visibly less accurate than for dense models.

Table~\ref{tab:trace-fidelity} summarizes these results across all model families, averaging over hardware and tensor-parallel settings. The results cleanly bifurcate: dense models, for which the compositional decomposition captures most of the observed dynamics, and MoE models, for which expert-routing we do not capture contributes a larger share of the remaining variation.

\begin{table}[t]
\centering
\small
\caption{Synthetic trace fidelity on held-out test data, averaged across hardware and TP configurations per model. Dense models use i.i.d.\ generation (Eq~\ref{eq:power-iid}), MoE models use AR(1) generation (Eq~\ref{eq:power-ar1}). Note, Llama-3.1 (405B) was only evaluated on an H100 with TP=8, therefore, there is no standard deviation to report.}
\label{tab:trace-fidelity}
\resizebox{\columnwidth}{!}{
\begin{tabular}{lcccc}
\toprule
\textbf{Model} &
\textbf{KS~$\downarrow$} &
\textbf{ACF $R^2$~$\uparrow$} &
\textbf{NRMSE~$\downarrow$} &
\textbf{$\Delta$Energy (\%)~$\downarrow$} \\
\midrule
Llama-3.1 (8B)         & 0.18~$\pm$~0.04           & 1.00~$\pm$~0.00           & 0.33~$\pm$~0.03           & 0.6~$\pm$~0.3        \\

Llama-3.1 (70B)             & 0.22~$\pm$~0.09           & 0.97~$\pm$~0.04           & 0.43~$\pm$~0.09           & 4.0~$\pm$~2.7        \\
Llama-3.1 (405B)           & 0.11                      & 0.97                      & 0.32                      & 4.5                  \\
DeepSeek-R1-Distill (8B)   & 0.21~$\pm$~0.08           & 0.99~$\pm$~0.01           & 0.32~$\pm$~0.05           & 1.0~$\pm$~0.4        \\
DeepSeek-R1-Distill (70B)  & 0.16~$\pm$~0.02           & 0.98~$\pm$~0.02           & 0.32~$\pm$~0.18           & 2.6~$\pm$~0.8        \\
\midrule
gpt-oss (120B)              & 0.51~$\pm$~0.11           & 0.58~$\pm$~0.07           & 0.33~$\pm$~0.04           & 10.8~$\pm$~3.5       \\
gpt-oss (20B)               & 0.48                      & 0.21                      & 0.31                      & 11.2                 \\
\bottomrule
\end{tabular}}
\end{table}

\begin{figure}[t]
    \centering
    \subfigure[Dense A100 TP4]{
    \includegraphics[width=0.31\columnwidth]{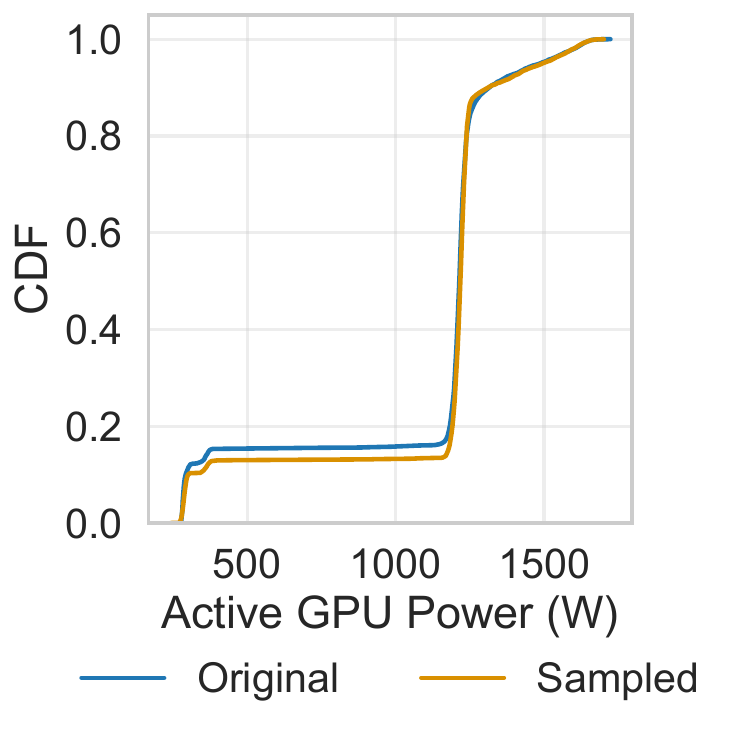}}
    \subfigure[Dense H100 TP1]{
    \includegraphics[width=0.31\columnwidth]{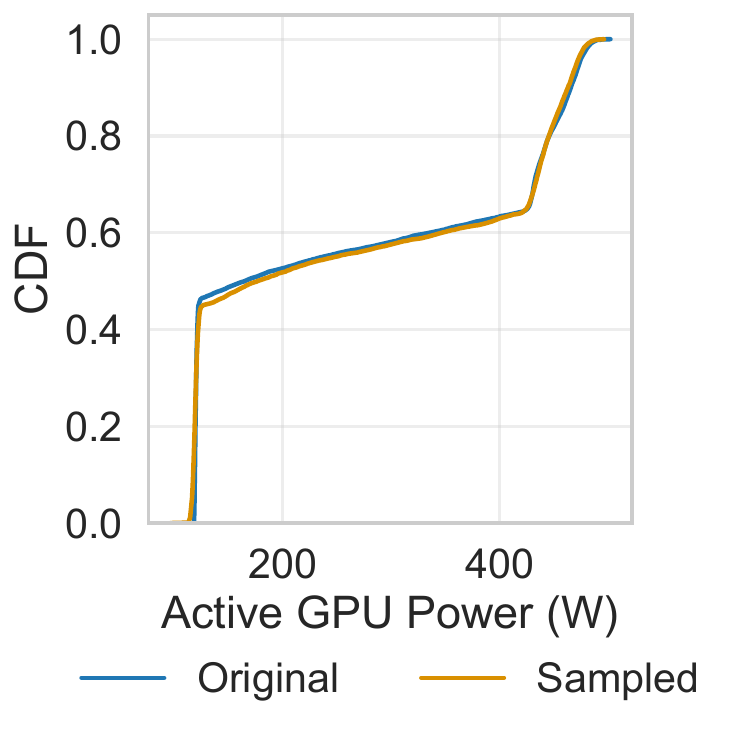}
    }
    \subfigure[MoE A100 TP8]{
    \includegraphics[width=0.31\columnwidth]{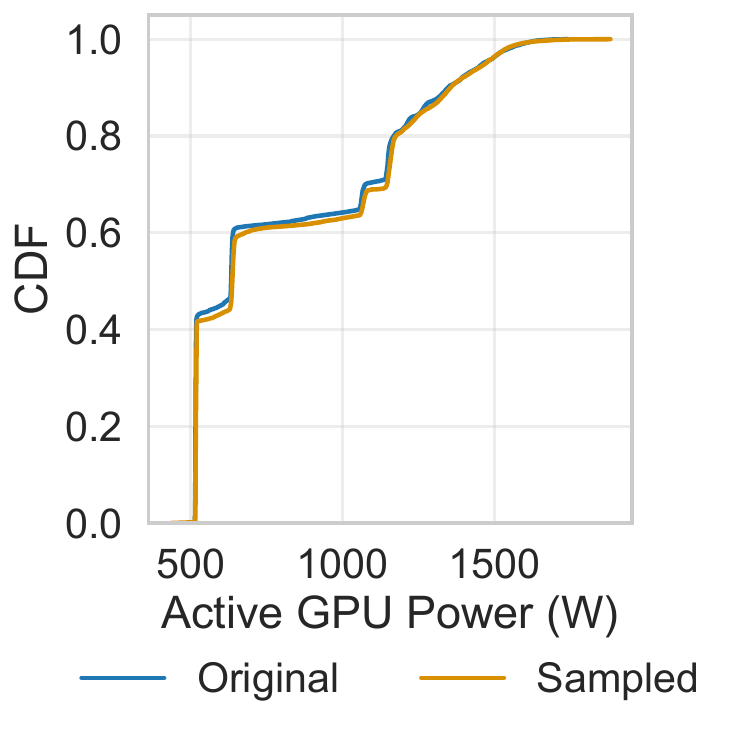}
    \label{fig:moe-cdf}
    }
    \caption{CDFs of synthetic versus measured power on held-out test data for representative dense and MoE configurations. In both cases, the marginal power distribution is reproduced more accurately than the full temporal dynamics. The models are (a) DeepSeek-R1-Distill (70B), (b) Llama-3.1 (8B), (c) gpt-oss (120B).}
    \label{fig:cdfs}
\end{figure}

\noindent\textbf{Dense models.}
All five dense configurations achieve strong temporal fidelity, with mean ACF~$R^2$ above 0.96 and median $|\Delta E|$ below 5\%. Figure~\ref{fig:cdfs} further shows that the synthetic traces reproduce the marginal power distributions of dense models well.

\noindent\textbf{MoE models.}
MoE fidelity is lower. The gpt-oss (120B) configuration achieves moderate temporal coherence (ACF~$R^2 = 0.58$) but relatively high energy error (10.8\%), while gpt-oss (20B) preserves energy well but exhibits little autocorrelation from our calculations. In both cases, expert routing causes within-state power variation that is not represented by $A_t$ and $\Delta A_t$, and the AR(1) correction captures only some low order persistence. Even so, marginal distributions remain well matched, as reflected in the KS statistics and in Figure~\ref{fig:moe-cdf}. This suggests that coarse energy and load-duration properties are able to be preserved by our models even when the exact temporal structure is not.

\subsection{Comparison with Simpler Power Models}
\label{sec:baselines}

We compare our framework against three baselines that represent some common methods used in planning studies.

\emph{TDP (nameplate)} assumes that every server draws rated TDP at all times. This is the most conservative abstraction and is often implicit in first-pass capacity planning and interconnection studies.

\emph{Mean power} assumes that every server draws its empirical training-set mean at all times,
$P_{\mathrm{server}}(t) = \bar{y}_{\mathrm{train}}$.
This captures the average operating point but removes all temporal variation.

\emph{Splitwise-style LUT baseline.}
We implement a Splitwise-inspired LUT baseline using the public Splitwise performance model and released characterization tables~\cite{patel2024splitwise}. The baseline predicts prefill and decode latency from batch token count and treats mixed iterations as prompt-like with a small penalty, and assigns phase-dependent power ratios for prompt, decode, mixed, and idle operation. Node power is then synthesized by scaling active-GPU power by the selected ratio from the LUT and adding fixed non-GPU overhead. Since the public Splitwise tables are calibrated on A100 and Llama-2 70B, whereas our experiments target Llama-3.1 70B at TP=8, we treat this baseline as a structurally matched LUT surrogate rather than an exact reproduction of the original system.

\noindent\textbf{Server-level comparison.}
Table~\ref{tab:baselines-node} reports the server-level metrics for representative dense configurations. There is no ACF~$R^2$ for TDP and mean power because they are constants.

\begin{table}[t]
\centering
\small
\caption{Baseline comparison at server level for Llama-3.1 (70B) A100 TP=4 and TP=8 data.}
\label{tab:baselines-node}
\begin{tabular}{lcccc}
\toprule
\textbf{Method} &
\textbf{KS~$\downarrow$} &
\textbf{ACF $R^2$~$\uparrow$} &
\textbf{NRMSE~$\downarrow$} &
\textbf{$\Delta$E (\%)~$\downarrow$} \\
\midrule
TDP & 1.00 & --- & 1.66 & 243.60 \\
Mean & 0.69 & --- & 0.32 & 17.35 \\
LUT-based & 0.64 & 0.56 & 0.27 & 13.71 \\
Ours & 0.12 & 0.99 & 0.27 & 6.09 \\
\bottomrule
\end{tabular}
\end{table}

TDP overestimates energy by more than $3.5\times$, reflecting the fact that inference GPUs spend much of their time below nameplate. Mean power removes most of that bias but, by design, cannot reproduce temporal structure. LUT-Based performs better than either constant baseline, but its three-level formulation is still too coarse. In our strict calibration, the phase levels miss the measured operating points entirely which results in poor distributional agreement, autocorrelation, and higher energy errors. The limiting factor is the abstraction, since this LUT approach cannot represent the occupancy of the GPU adequately to compute dependent power that appears when concurrent requests are filling and draining the batch. Our framework can recover that structure as shown by lower energy error and high temporal correlation.

\begin{figure}[t]
    \centering
    \includegraphics[width=\columnwidth]{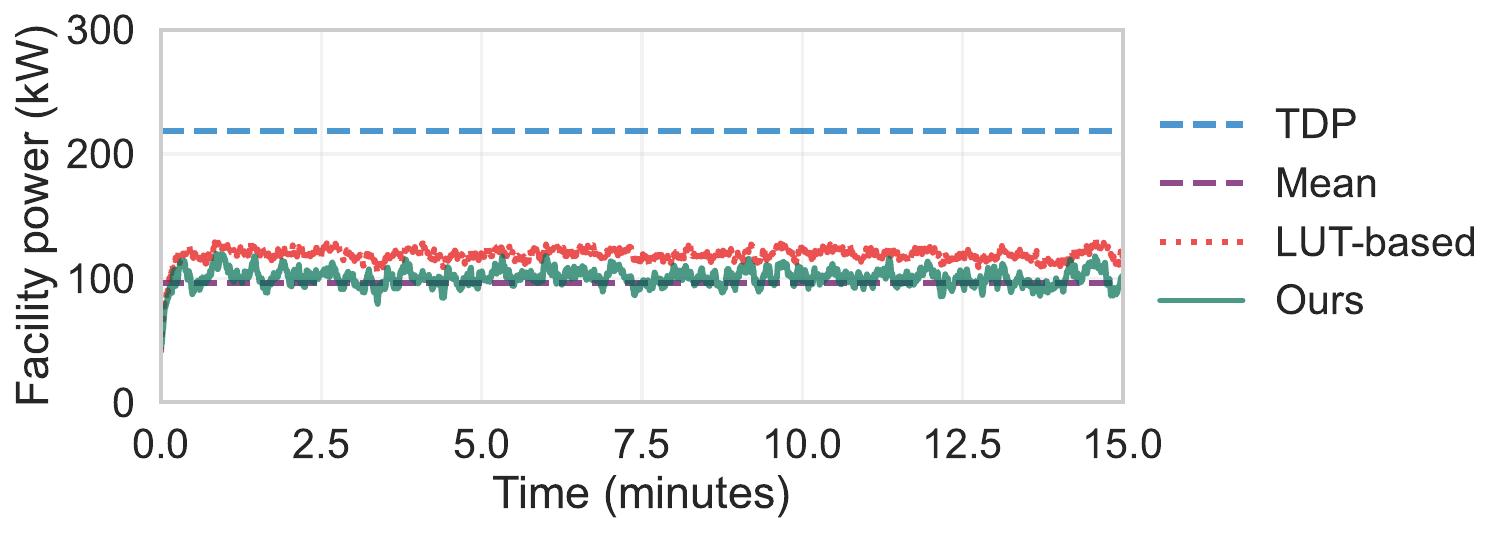}
    \caption{15 minutes of facility power (60 servers, Llama-3.1 (70B), H100, TP$=$8) generated by each method.}
    \label{fig:baseline-facility}
\end{figure}

\noindent\textbf{Facility-level comparison.}
These differences are still visible after aggregation. Figure~\ref{fig:baseline-facility} shows 15 minutes of facility power for a 60-server deployment (Llama-3.1 (70B), H100, TP$=$4) under Poisson arrivals. LUT-Based performs worse still, missing both the overall magnitude and the temporal pattern. 

\subsection{Infrastructure Planning with Production Workloads}
\label{ssec:facility}

We now use a production workload to show what our traces reveal about infrastructure planning decisions. We drive our pipeline with a production-representative Azure LLM inference trace~\cite{dynamollm} that exhibits both a diurnal structure and bursty arrivals. We then study three planning views of the same facility: facility load profiling, interconnection sizing, and rack-level oversubscription.

\noindent\textbf{Facility configuration.}
We model an arbitrary data hall with $10$ rows, $6$ racks per row, and $4$ servers per rack, for a total of $240$ servers. We assume that all servers run Llama-3.1 (70B) on A100 racks with TP$=8$. Each server receives its own request stream derived from the Azure trace with a random temporal offset so that arrivals are decorrelated across the facility. We assume 1\,kW of non-GPU IT power per server and PUE$=1.3$, and generate 24 hours of facility power at 250\,ms resolution.

\begin{figure}[t]
    \centering
    \includegraphics[width=\columnwidth]{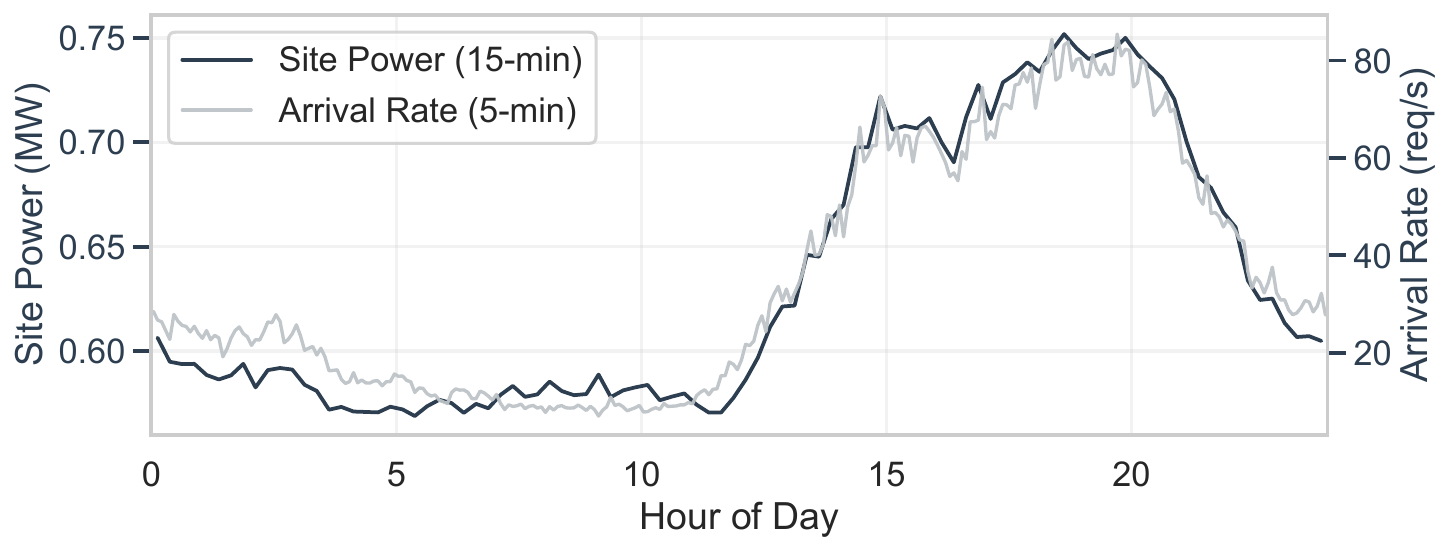}
    \caption{24-hour facility power profile (240 servers, PUE$=1.3$) driven by a production Azure inference trace~\cite{dynamollm}. Solid line: 15-minute site power. Dashed line: 5-minute arrival rate.}
    \label{fig:facility-24hr}
\end{figure}

Figure~\ref{fig:facility-24hr} shows the resulting 24-hour facility profile at a 15-minute resolution, together with the input arrival rate. This is the kind of load shape that would be used in an interconnection or substation study. The site power follows the same broad diurnal pattern as the workload, but not a scaled copy of the arrival trace. The mapping from arrivals to power is nonlinear, and bursts can also introduce lag because requests queue before they reach the GPU. Peak power occurs during the afternoon surge, while overnight power drops toward idle but still reflects the variability of sparse inference traffic.

\noindent\textbf{Interconnection sizing.}
Table~\ref{tab:facility-sizing} summarizes quantities that can be extracted from this 24-hour facility trace.

\begin{table}[t]
\centering
\small
\caption{Infrastructure sizing from 24-hour facility simulation (240 servers, PUE\,$=$\,1.3).}
\label{tab:facility-sizing}
\begin{tabular}{lccccc}
\toprule
\textbf{Metric} & \textbf{TDP} & \textbf{Mean} & \textbf{LUT-Based} & \textbf{Ours} \\
\midrule
Peak facility power (MW) & 1.19 & 0.85 & 0.82 & 0.75 \\
Average facility power (MW) & 1.19 & 0.85 & 0.76 & 0.63 \\
Peak-to-average ratio & 1.00 & 1.00 & 1.09 & 1.19 \\
Max ramp rate (MW/15-min) & 0.00 & 0.00 & 0.07 & 0.11 \\
Load factor & 1.00 & 1.00 & 0.92 & 0.84 \\
\bottomrule
\end{tabular}
\end{table}

TDP would size the facility at 1.19\,MW, while our generated traces produce a 24-hour peak of 0.75\,MW. In this case, nameplate provisioning overstates required interconnection capacity by roughly 60\%. Mean power captures only the long-run operating point, which removes all variability from the profile. It also underestimates the facility average, since the token trace sustains a higher operating point than the uniform arrival-rate sweep used during training. LUT-Based overstates this slightly at 0.82\,MW and underestimates the ramping effects that are present from our approach, a critical quantity for grid operators to evaluate a site. Our power traces instead produce a peak-to-average ratio of 1.19, a load factor of 0.84, and a maximum 15-minute ramp of 0.11\,MW. These results illustrate quantities that matter in planning, capacity allocation, and downstream grid studies. 

\noindent\textbf{Oversubscription analysis.}
We next consider planning inside the facility, where row-level power constraints determine how much IT load can be deployed safely. For this example, we assume a 600\,kW row distribution limit and require that the 24-hour peak row power remain below that limit across 5 random seeds.

\begin{figure}[t]
    \centering
    \includegraphics[width=0.85\columnwidth]{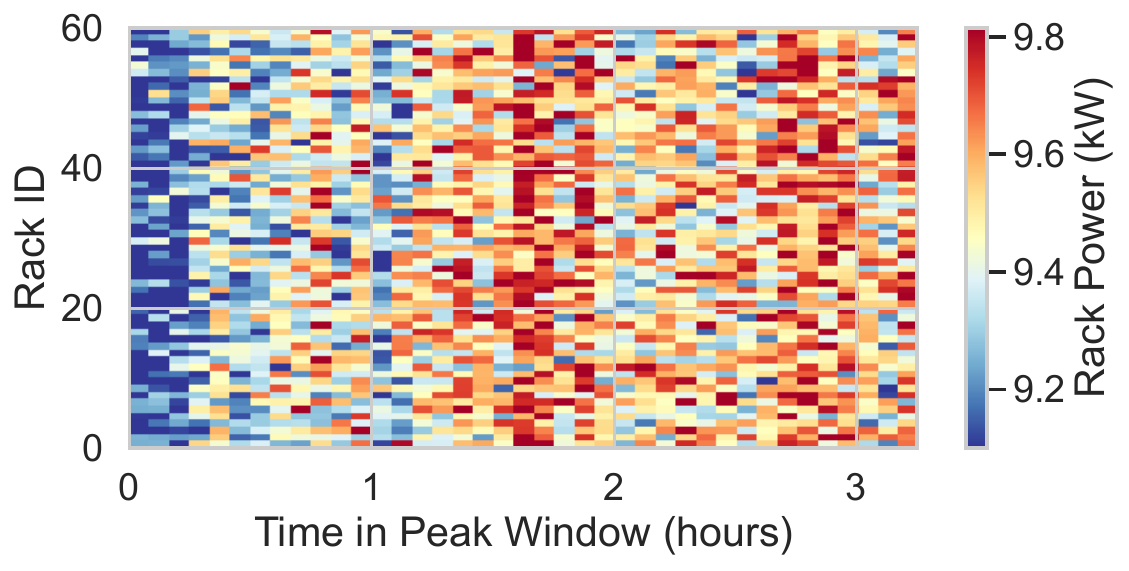}
    \caption{Per-rack power heatmap during a four-hour peak window, showing temporal decorrelation across racks serving independent Azure inference streams.}
    \label{fig:heatmap}
\end{figure}

Figure~\ref{fig:heatmap} shows per-rack power over a four-hour peak window. The main pattern is that there is temporal decorrelation across racks, since under independent arrivals, individual rack peaks do not align. This makes the row-level aggregate remain much smoother than the per-rack traces. TDP cannot represent this because it assumes permanent full utilization, and mean power cannot represent it because it assumes no peaks at all.

\begin{figure}[t]
    \centering
    \includegraphics[width=0.9\columnwidth]{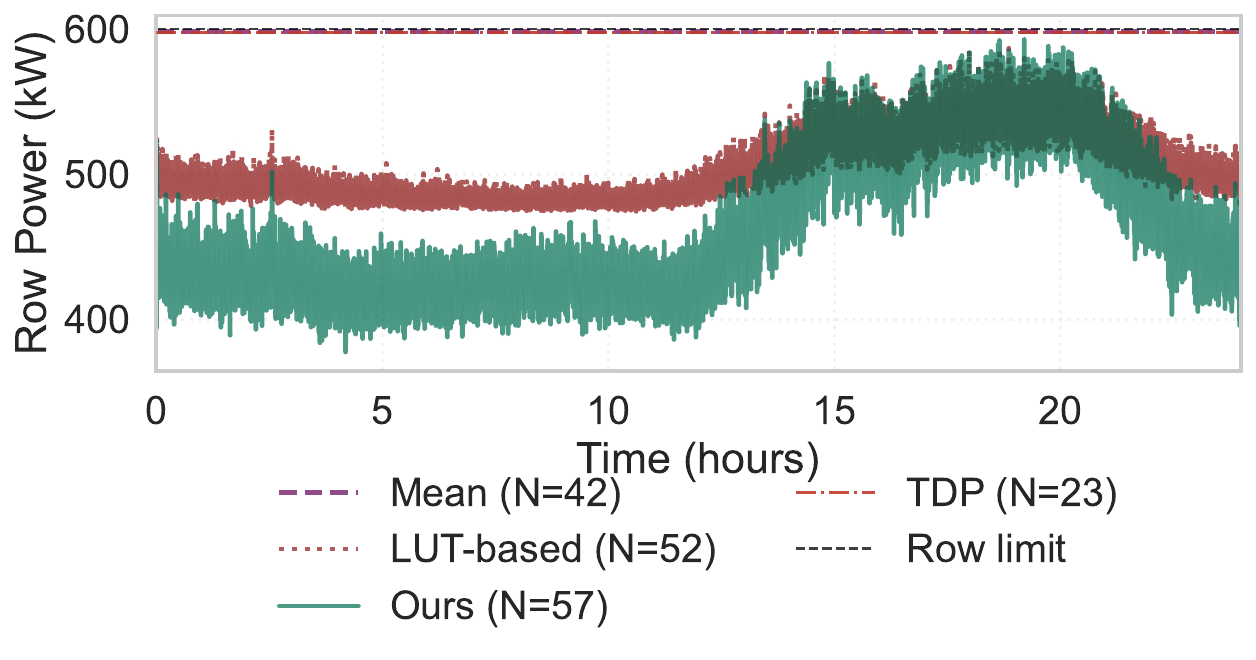}
    \caption{Aggregate row power for deploying racks above the TDP recommended nameplate capacity.}
    \label{fig:oversub-row-power}
\end{figure}

Figure~\ref{fig:oversub-row-power} quantifies the resulting headroom. Under nameplate provisioning, a 600\,kW row can host only
$\lfloor 600\,\mathrm{kW}/\mathrm{rack\ TDP} \rfloor = 23$
racks. Under the production workload, however, those 23 racks reach only about 250\,kW at peak, leaving roughly 350\,kW of row capacity unused. To find the maximum number of racks deployable under this limit, we provision 4$\times$A100 DGX server racks running similar workloads until the P95 of the row power exceeds the 600\,kW limit. At this point we consider the row saturated. We find that using our power traces, this same row can accommodate 57 racks while remaining below the 600\,kW limit, with an observed peak of about 580\,kW. This corresponds to more than double the rack density of TDP-based provisioning in this case study. LUT-Based and Mean racks show similar deployments, reaching 52 and 42 racks, respectively, under the same deployment criteria.

We should note that this result depends on arrival-correlation assumptions. While we are using production token arrival data, if arrivals were more strongly synchronized across racks, row peaks could coincide more often and the oversubscription limit would fall. In practice, this means that oversubscription is evaluated as a function of traffic correlation rather than from a single static trace. Even so, this example shows the basic planning value of our framework, as it can expose headroom that static nameplate assumptions leave unused while still making the dependence on workload structure explicit.

\subsection{Aggregation Across the Hierarchy}
\label{ssec:aggregation}

The facility-scale results above rely on the simple property of aggregation, that as independent server-level traces are summed through the datacenter hierarchy, short-timescale variability is smoothed.

\begin{figure}[t]
    \centering
    \subfigure[Server]{
    \includegraphics[width=0.47\columnwidth]{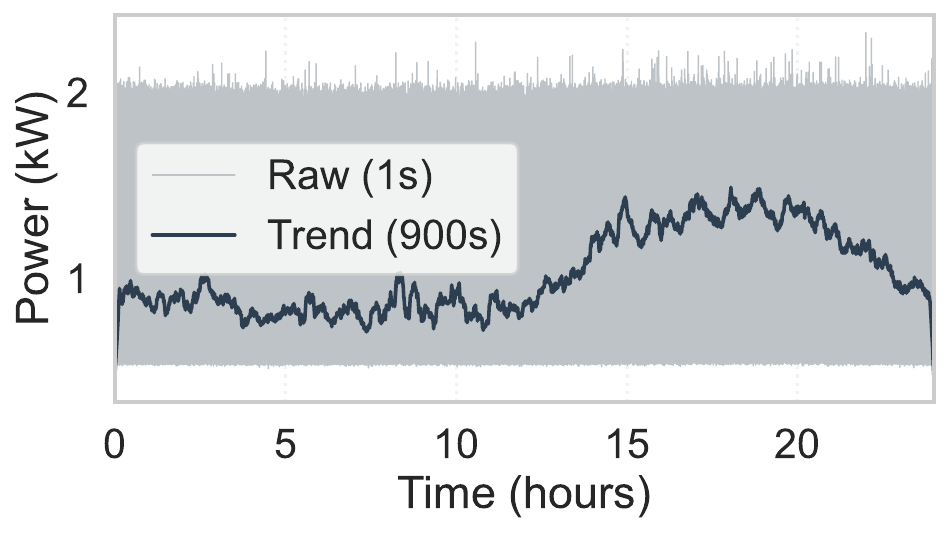}
    }
    \subfigure[Rack]{
    \includegraphics[width=0.47\columnwidth]{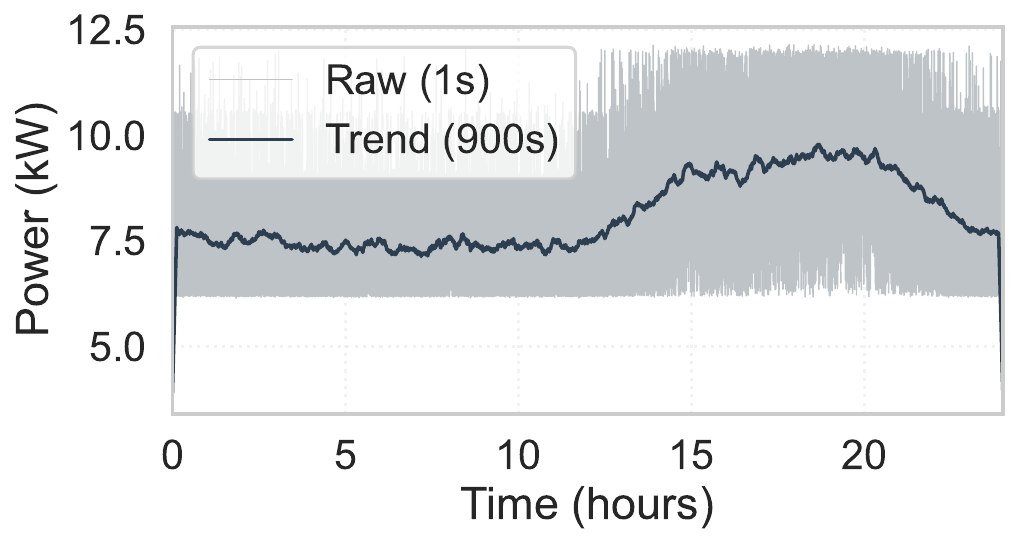}
    }
    \subfigure[Row]{
    \includegraphics[width=0.47\columnwidth]{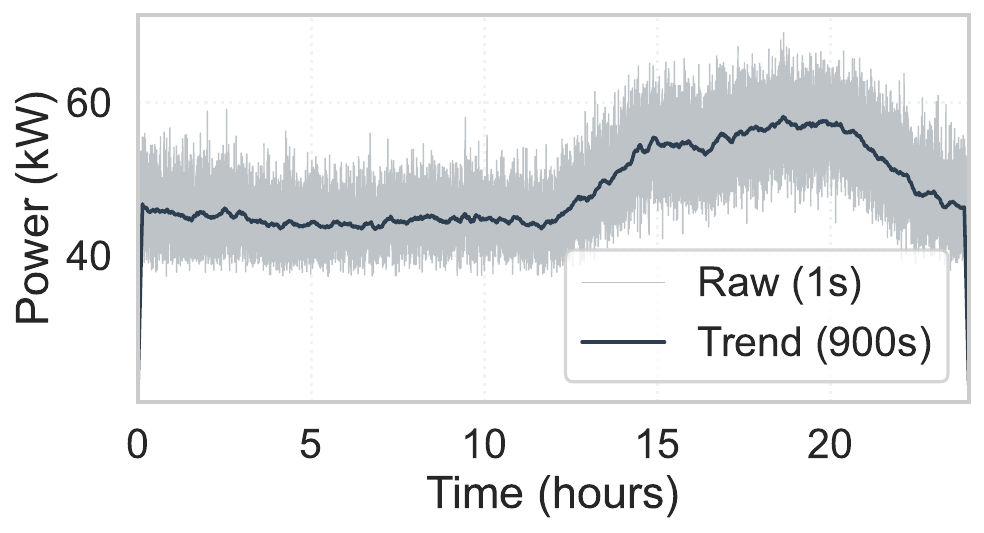}%
    }
    \subfigure[Site]{
    \includegraphics[width=0.47\columnwidth]{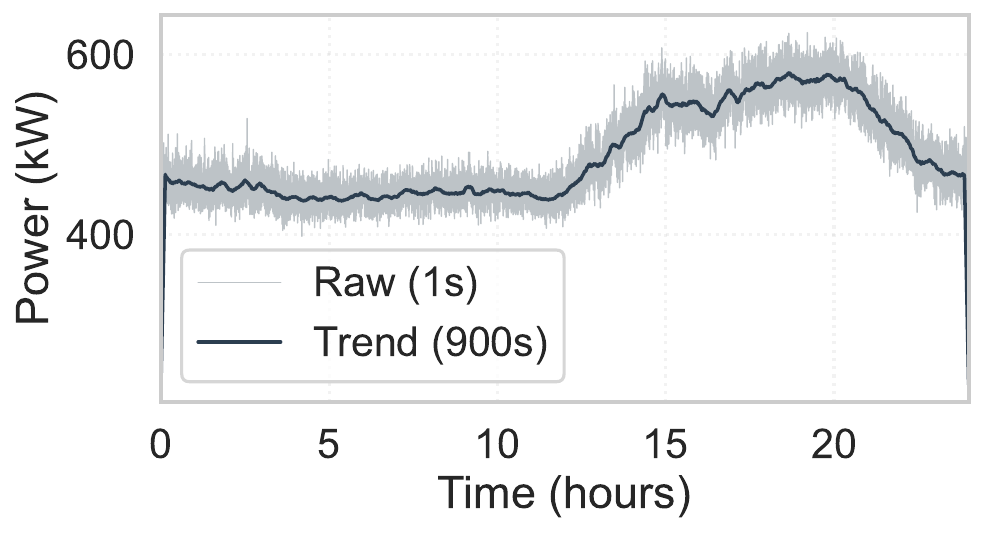}}
    \caption{Power traces at four hierarchical levels over 24 hours. As you aggregate, variance decreases and trends towards the mean.}
    \label{fig:hierarchy}
\end{figure}

Figure~\ref{fig:hierarchy} shows this progression over the same 24-hour period. At the single-server level (250\,ms), power exhibits sharp transitions characteristic of inference serving, spanning nearly the full range from idle to near-TDP. Aggregating four servers into a rack already compresses that and softens some of these transitions. At the row level, transient excursions are further averaged out, and the load fluctuates within a much narrower band. By the site level, viewed at 15-minute resolution, the dominant signal is the diurnal envelope that is visible from the utility.

This smoothing matters in two ways. First, it is the mechanism that creates oversubscription headroom. Server-level peaks do not coincide perfectly, so row and site level demand remains below the sum of peak utilization. In our case study, the coefficient of variation falls from 0.583 at the single-server level to 0.127 at the site level, consistent with the expected reduction in variability under aggregation of largely independent sources. More strongly correlated arrivals, as seen in production~\cite{wang2024burstgpt,xiang2025servegen}, would reduce this benefit.

Second, aggregation makes facility-level planning less sensitive to small server-level errors, but not to systematic modeling bias. Random mismatch in exact transition timing or within-state variation can average out across many servers, but structural errors can distort aggregate peaks, ramps, and coincidence consistently. This is why aggregation does not rescue simpler approaches even though it can reduce the effect of residual error in our own traces.

\section{Discussion}
\label{sec:discussion}

Our results suggest both a practical role for synthetic inference traces in planning workflows and a clearer view of the assumptions under which those traces can remain reliable. Here, we discuss these implications, along with the main limits of the current framework.

\subsection{Utility-Operator Coordination}
\label{ssec:privacy}

Our framework enables utility-facing load characterization for AI facilities before deployment. This matters because interconnection and planning studies increasingly require more than a flat nameplate value or annual energy forecast~\cite{norris2025rethinking, load-planning-reliability-considerations, capacityplanning,lin2024exploding}, while operators have strong incentives not to reveal serving details such as traffic composition, batching policy, or internal telemetry. 

The framework also enables a counterfactual analysis that is not possible from one-off forecasts. A utility can stress-test a proposed deployment under alternative assumptions about traffic intensity, model mix, or deployment scale, even before the facility is built. This better matches the probabilistic structure of existing planning workflows~\cite{capacityplanning}.

The emergence of closed-source platforms such as Emerald AI's Conductor~\cite{emeraldai} further illustrates the value of our interface. The distinction is that our framework is open-source and can be run by the operator, so synthetic traces or aggregate planning artifacts can be shared externally while raw serving telemetry remains internal. We do not claim formal privacy guarantees; the contribution is a practical reduction in disclosure surface, not a cryptographic one.

\subsection{Composability and Extensions}
\label{ssec:extensions}

The evaluation in this paper focuses on fixed serving configurations under arbitrary arrivals. The same compositional structure supports several natural extensions.

\emph{Model mix evolution and hardware refresh.} Because power behavior is modeled per per-configuration, adding a new model or accelerator does not require rebuilding the entire pipeline. A small set of measurements and retraining is sufficient, after which mixed deployments can be simulated directly. This makes the framework useful for studying phased rollouts, hardware refreshes, and partial migrations before they are deployed at scale.

\emph{Agentic and multi-step workloads.} Tool-calling agents and multi-turn reasoning systems introduce self-correlated arrivals, where one completion can trigger additional requests~\cite{li2026continuumefficientrobustmultiturn, liu2026droidspeak, kulkarni2026optimizingfaasplatformsmcpenabled, yao2022react}. In our framework, this primarily changes the arrival process rather than the per-server power model itself. If an agentic simulator or production trace produces a request schedule, then the same pipeline can translate that schedule into power traces. What becomes more important in this setting is cross-server correlation, since agentic bursts may reduce the diversity that makes oversubscription feasible.

\emph{Long-horizon planning and grid interaction.} Utilities and planners often require annual load shapes for resource-adequacy and capacity-expansion studies~\cite{gridstrategies2025,nerc2026reliability,load-planning-reliability-considerations,AESO2025datacentreconnection}. The framework can generate these from year-long arrival scenarios that reflect expected traffic growth, model turnover, or hardware replacement. The same traces can also be used to study potential load flexibility. If operators can reshape some portion of inference demand through queueing, routing, or deferral of latency-tolerant work, then scenario-generated traces can quantify when and how much reduction is actually available under deployment-specific SLOs.

\subsection{Scope and Limitations}
\label{sec:limitations}

\emph{Arrival processes.} Server-level fidelity is validated under Poisson arrivals. The Azure facility study demonstrates plausible behavior under production-like diurnal and bursty demand, but we do not claim model adherence across all distributions of arrivals.

\emph{Facility modeling.} We use a constant PUE model and constant non-GPU IT overhead. This is appropriate for the planning-scale comparisons in the paper, but it does not capture weather-dependent cooling behavior, load-dependent PUE variation, or auxiliary systems whose power may vary over time~\cite{EPRI2024PoweringDataCenters}.

\emph{Serving policies and hidden internal dynamics.} The throughput model abstracts away scheduler internals such as memory-aware batching, preemption, and other serving-engine details. Our classifiers are trained on vLLM~\cite{vllm}, so configurations that materially change the implicit mapping from arrivals to kernel usage and hence power may require some retraining. More broadly, the framework is strongest when workload-visible features explain most of the trace dynamics. This is true for the dense-model settings we study, but not as much for MoE systems, where the hidden expert-routing adds some within-state variation that our arrival-level features can only partially capture.

\section{Related Work}
\label{sec:related}

\noindent\textbf{Grid-facing load construction for datacenters.} Some approaches exist for constructing datacenter load profiles for grid and planning studies. In practice, utilities often rely on customer-provided profiles when evaluating new large loads, and regulatory processes increasingly require projected load shapes as part of the interconnection process itself~\cite{capacityplanning, norris2025rethinking, load-planning-reliability-considerations, AESO2025datacentreconnection}. Measurement-based forecasting requires an already-operational facility~\cite{mughees2025shorttermloadforecastingaidata}, while scenario studies often move away from the serving configuration that drives short-timescale power variation~\cite{chen2025electricitydemandgridimpacts, vabson2026optimalcountylevelsitingdata}. We instead provide traces before full deployment, rather than after a facility is already operating.

\noindent\textbf{LLM inference power characterization and control.} A growing body of work shows that inference power has recognizable structure, especially around the prefill--decode distinction, and that this structure can be exploited both architecturally and operationally. On the architectural side, phase separation has been used to motivate disaggregated serving, energy-aware placement, and power-aware reallocation across serving pools~\cite{patel2024splitwise, distserve, basit2026biscaleenergyefficientdisaggregatedllm, jiang2026rapid}. On the control side, phase-aware energy management and finer-grained power control have been studied under bounded SLO or SLA constraints~\cite{stojkovic2025tapas, dynamollm, polca, muserve, kakolyris2025throttlem, spaan2026reducingcomputewastellms,emeraldai}. Measurement studies further quantify how batching, quantization, and hardware configuration shape inference energy~\cite{delavande2026understandingefficiencyquantizationbatching, koomey2024single,fernandez2025energy,wilkins2024hybrid,wilkins2024offline}. This literature shows that inference power is structured and controllable, something we use to generate traces for new scenarios.

\noindent\textbf{Datacenter power modeling and oversubscription.} Datacenter power management spans device-level profiling and simulation on one end~\cite{kandiah2021accelwattch, you2023zeus, chae2024perseus,joulemeter} and facility-scale demand management and oversubscription on the other~\cite{wu2016dynamo, radovanovic2022modeling, mvplane-powercapping, zhang2021flex, harvesting, hsu2018smoothoperator, reidys2025coach, piga2024dvfs,pelley2010power}. Our work is complementary, providing the workload-conditioned inference traces these systems and studies would consume.

\noindent\textbf{Workload generation for AI inference.} The closest prior work studies synthetic workload generation for cloud and AI serving systems. BurstGPT~\cite{wang2024burstgpt} and ServeGen~\cite{xiang2025servegen} generate production-representative request streams, but stop at arrivals and do not model power. Bergsma et al.~\cite{bergsma2021complex} show that recurrent sequence models can capture long-range temporal structure in cloud workloads, which is closely aligned with our use of sequence modeling, but their target is workload generation rather than power synthesis. More general time-series generation methods~\cite{rousseau2025forging} produce synthetic sequences without modeling the state-structured serving dynamics that govern inference power, and inference simulators likewise stop short of calibrated, deployment-grounded power traces~\cite{agrawal2024etalonholisticperformanceevaluation, patel2024splitwise}. These tools model the demand side of inference workloads; our focus is the resulting power they induce.

\section{Conclusion}

In this paper, we presented a compositional framework for generating LLM inference power traces from workload scenarios and limited server-level measurements. By separating workload-driven state evolution from configuration-specific power behavior, the framework can synthesize realistic traces across traffic conditions, hardware platforms, and serving settings without end-to-end profiling of each new scenario. Across the settings we profile, our model reproduces measured energy and temporal structure with high fidelity. Our case studies show that this additional accuracy matters, since flat TDP and mean-power abstractions miss the peaks, ramps, and coincidence structure that determine provisioning outcomes and oversubscription headroom. Overall, this work provides a practical path from inference-serving behavior to the planner-facing electrical load models needed for datacenter infrastructure design.

\begin{acks}
Grant Wilkins was supported in part by a Stanford University Bits \& Watts Initiative Powering AI Seed Grant. We thank Microsoft Azure Research -- Systems for providing access to the compute nodes used for profiling and data collection in this project.
\end{acks}

\bibliography{main}
\bibliographystyle{ACM-Reference-Format}

\appendix
\onecolumn
\section{Workload-Feature Adherence}

\subsection{Surrogate Reproduces Concurrency Dynamics}\label{app:surrogate}

\begin{figure}
    \centering
    \subfigure[Deepseek-R1-Distill (70B) A100 TP8, $\lambda=4.0$]{
    \includegraphics[width=0.35\columnwidth]{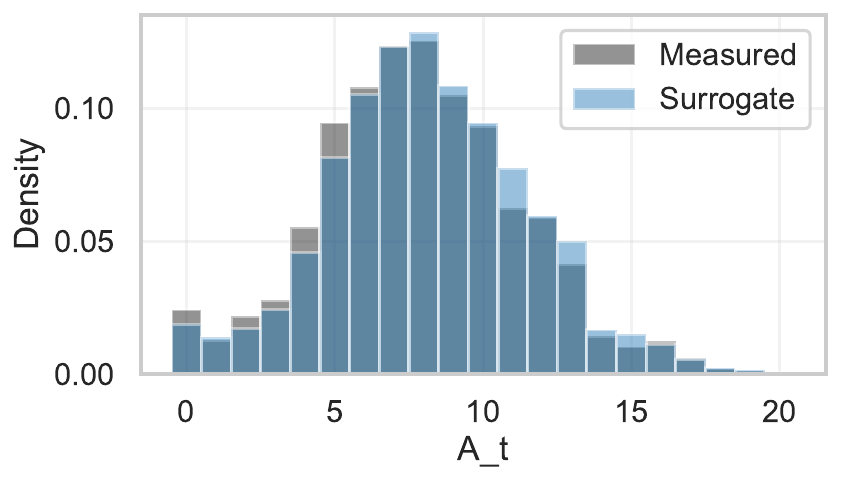}}
    \subfigure[Deepseek-R1-Distill (70B) A100 TP8, $\lambda=4.0$]{
    \includegraphics[width=0.35\columnwidth]{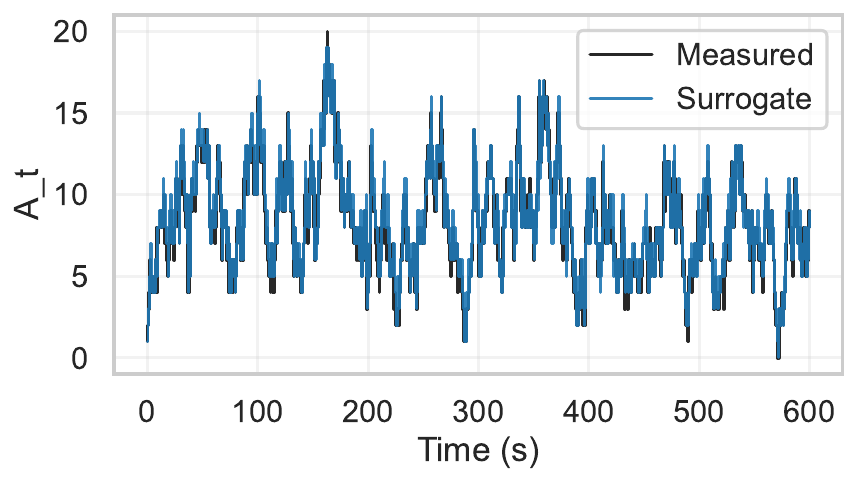}}
    \subfigure[Deepseek-R1-Distill (70B) H100 TP4, $\lambda=0.25$]{
    \includegraphics[width=0.35\columnwidth]{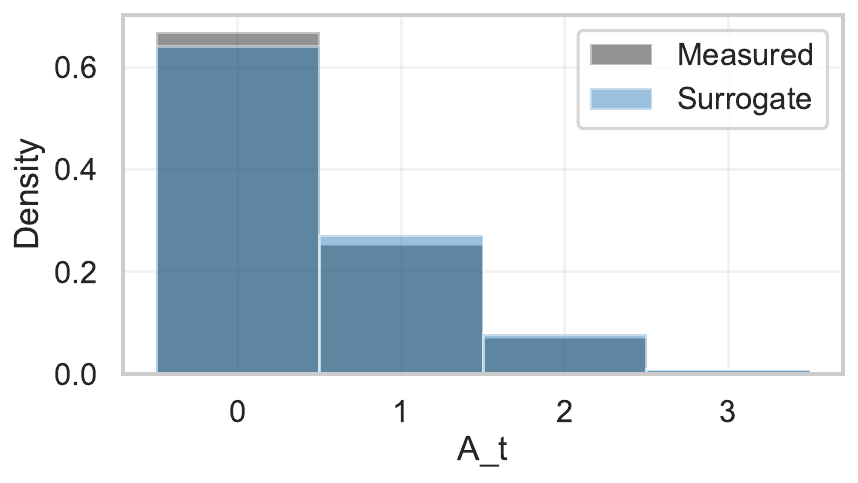}}
    \subfigure[Deepseek-R1-Distill (70B) A100 TP4, $\lambda=0.25$]{
    \includegraphics[width=0.35\columnwidth]{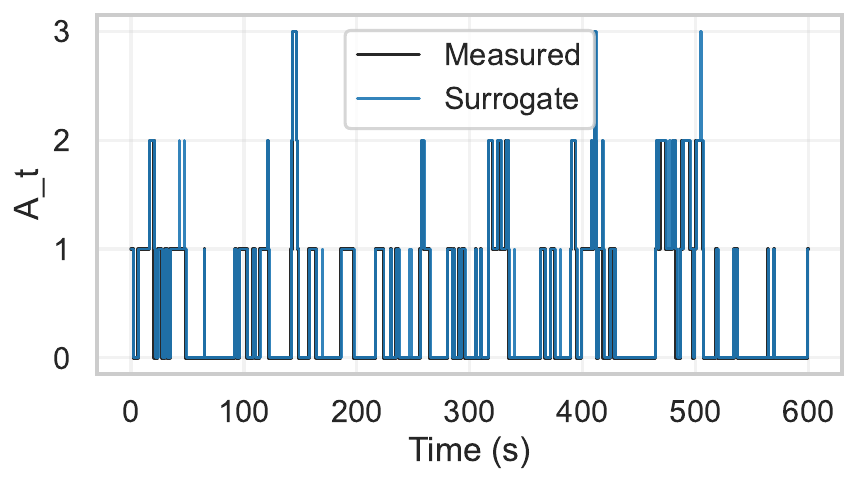}}
        \subfigure[Deepseek-R1-Distill (70B) H100 TP8, $\lambda=0.5$]{
    \includegraphics[width=0.35\columnwidth]{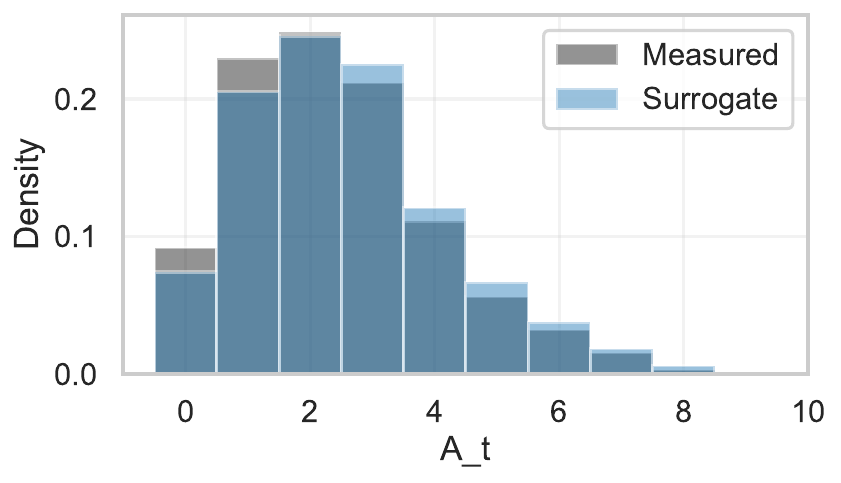}}
    \subfigure[Deepseek-R1-Distill (70B) H100 TP8, $\lambda=0.5$]{
    \includegraphics[width=0.35\columnwidth]{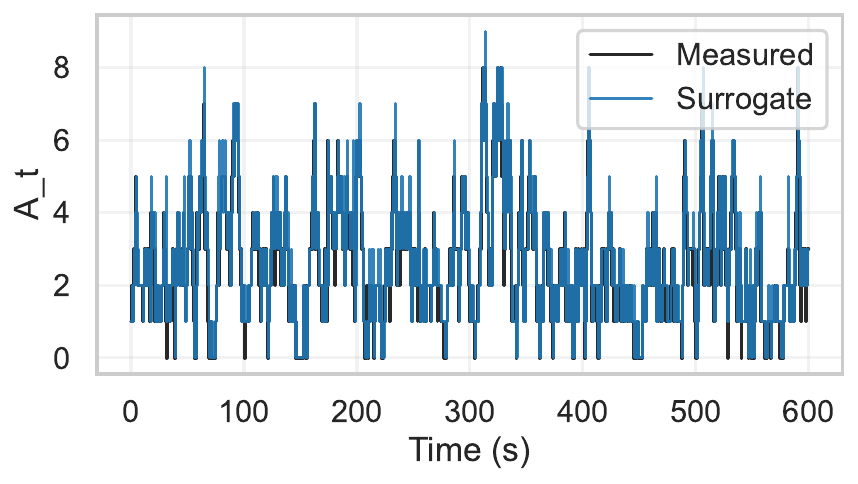}}
    \caption{$A_t$ reproductions from fit token latency surrogate and FIFO queue model.}
    \label{fig:arrival-reproduction}
\end{figure}

Our framework decouples power modeling from the serving-engine by computing workload features $(A_t, \Delta A_t)$ from a throughput model rather than from runtime instrumentation (\S3.3). This surrogate approximates the serving stack as a FIFO queue with log-linear prefill latency (Eq.~\ref{eq:ttft}) and lognormal decode latency (Eq.~\ref{eq:tbt}). If the composed system fails to reproduce realistic concurrency dynamics, the downstream classifier receives out-of-distribution inputs and trace fidelity will degrade regardless of model quality. Here in Figure~\ref{fig:arrival-reproduction} we show the throughput against measured $A_t$ trajectories for DeepSeek-R1-Distill (70B) across two GPU generations, two tensor-parallel settings, and three arrival rates spanning the low-to-high load regime. At high load ($\lambda = 4.0$ req/s), the model can capture both the mode and spread of the concurrency distribution. At low load ($\lambda = 0.25$ req/s), both measured and surrogate $A_t$ concentrate on 0--1 active requests with some excursions, reproducing the telegraph-like arrival pattern that drives sharp, idle-to-active power transitions. At moderate load ($\lambda = 0.5$ req/s), the surrogate matches the intermediate occupancy regime where multiple requests overlap but the GPU has not yet saturated. This surrogate is not intended to reproduce exact request-level timing, but rather to produce $A_t$ trajectories whose distributional and temporal properties align enough to the classifier's training distribution. The correlation and distributional agreement in Figure~\ref{fig:arrival-reproduction} shows that the features a capacity planner computes from an arrival schedule and a small latency calibration are realistic enough to drive the learned power model.

\end{document}